# Ambipolar Graphene Field Effect Transistors by Local Metal Side Gates

J. F. Tian[*, a, b], L. A. Jauregui[c, b], G. Lopez[c, b], H. Cao[a, b], and Y. P. Chen[*, a, b, c]

[a]Department of Physics, Purdue University, West Lafayette, Indiana 47907, USA

[b]Birck Nanotechnology Center, Purdue University, West Lafayette, Indiana 47907, USA

[c]School of Electrical and Computer Engineering, Purdue University, West Lafayette, Indiana 47907, USA

## Abstract

We demonstrate ambipolar graphene field effect transistors individually controlled by local metal side gates. The side gated field effect can have on/off ratio comparable with that of the global back gate, and can be tuned in a large range by the back gate and/or a second side gate. We also find that the side gated field effect is significantly stronger by electrically floating the back gate compared to grounding the back gate, consistent with the finding from electrostatic simulation.

[*] E-mails: tian5@purdue.edu, yongchen@purdue.edu

Graphene, composed of a two-dimensional (2D) hexagonal carbon lattice, stands out as a potential candidate for nanoelectronics and devices applications.[1-3] Its unusual band structure has a linear energy-momentum relation near the Dirac point where the valence and conduction bands meet, making graphene a zero-gap semiconductor. Both the type (electron or hole) and density of carriers in graphene can be easily controlled by using an electric field. Such an ambipolar electric field effect underlies a large number of work on graphene studying its electronic transport[4-7] as well as sensing[8,9] and other device-related applications[10-13]. The simplest and most common graphene field effect transistors (GFET) employ a heavily doped Si substrate as a global back gate. Such a global back gate tunes all the devices on the same substrate. In order to control individual GFETs (required in integrated circuits) and to realize more complex graphene devices, local gating using either top gates or side gates is required.[14-18] For example, top gates have been used to realize graphene-based bipolar devices (such as p-n junctions) within a single sheet.[14,16,17] However, deposition of dielectric required in the top gate fabrication could potentially lead to degradation of graphene mobility, and care must be taken and only very recently progresses have been made to overcome the detrimental effect.[19-23] Molitor *et al.* and Li *et al.* reported using graphene as side gates fabricated by reactive ion[15] and oxygen plasma etching[18], respectively. While capable to realize lateral modulation of charge density, this approach is limited by the size of graphene and the etching process can also degrade the graphene quality. Therefore, developing an easy and "clean" way of local gating on graphene is still desirable. Graf *et al.* gated a mesoscopic graphite wire[24] using metal side gates, with only limited efficacy due to the relative thickness of the graphite. In this work, we demonstrate ambipolar graphene field effect transistors (GFET) individually gated by local metal side gates, and investigate and analyze the widely tunable field effect transport in such graphene nanodevices.

Our graphene samples are prepared by micromechanical exfoliation of graphite on top of 300 nm $SiO_2$ on heavily doped p-type ($p^{++}$) Si substrate.[25] Monolayer graphene can be identified by its optical contrast[25] and distinctive Raman spectrum[26]. Metal side gates and contacts electrodes (5nm Ti+30nm Au) are fabricated in a one-step process using e-beam lithography, metal deposition, and lift-off. No gate dielectric deposition or etching of graphene is needed in the process. Figure 1a shows the atomic force microscopy (AFM) image of a representative GFET device with two metal side gates (device "1"). The typical distance between the side gate electrodes and edge of graphene ranges from several tens to hundreds of nanometers and is ~370nm for this device. A smaller distance gives stronger capacitive coupling between the side gate and graphene. Figure 1b shows the 3D schematic structure of the device and the corresponding circuits used in the measurements. All resistance measurements (4-terminals, see Fig. 1b) are performed at room temperature and in vacuum (<6 mTorr) by lock-in detection with a driving current of 100 nA. All the gate voltages are applied by DC source meters. Figure 1d shows the room temperature back gate field effect for device "1". A characteristic ambipolar field effect is observed with a global Dirac point ($V_{DP}$) at 24V. The positive $V_{DP}$ indicates p-type doping, probably due to polymethyl methacrylate (PMMA) residue or adsorption of molecules (such as water) on the graphene surface. Our fabricated graphene devices have typical carrier mobilities of ~2000-5000 $cm^2$/Vs extracted from both Hall effect and back-gated field effect measurements.

We have investigated the field effect controlled by both side and back gates. Data measured in device "1" are presented in Fig. 2. For simplicity, only one of the side gates (SG1, Fig. 1a) is used (using SG2 gives similar results). Figure 2a shows the resistance (R) as a function of the side gate voltage ($V_{sg}$) at a series of back gate voltages ($V_{bg}$) varied from 14.2V to 25.4V with a step of 0.8V. When the $V_{bg}$ =14.2V (much lower than $V_{DP}$ =24V, and the entire graphene being heavily p-type), R increases with the increasing $V_{sg}$ within the measurement range. When the $V_{bg}$ ~15.8V, a clear "side gate Dirac point" ($V_{SDP}$) with maximal R appears around 40V. Upon further increasing $V_{bg}$, $V_{SDP}$ decreases (from positive toward negative) while the maximal R of the device increases till $V_{bg}$ reaches ~21.4V then decreases again. These results show that the side-gated field effect can be tuned by the back gate. We have also studied tuning of the back gated field effect for this device by the side gate, by measuring R as a function of $V_{bg}$ at various $V_{sg}$ (Fig. 2b). We observe similar as in Fig. 2a, that is, the back-gate charge neutral point ($V_{DP}$) of the device continuously decreases as $V_{sg}$ increases (from -60 to 80V), while the R maximum in the $V_{bg}$ sweep first increases then decreases (reaching a peak at $V_{sg}$ ~7.5V). Figure 2c shows the 2D color-scale plot of R as a function of both $V_{bg}$ and $V_{sg}$. The vertical and horizontal cuts in such a color plot correspond to similar side-gated and back-gated field effect curves shown in Fig. 2a and 2b respectively. The color plot (Fig. 2c) clearly shows the shift of side/back gated field effect as controlled by the back/side gate, as well as a global maximum in resistance as $V_{sg}$ ~0V and $V_{bg}$ ~25V. Qualitatively similar tuning of back and side gated field effect were previously demonstrated with graphene side gates.[15]

We have also investigated the field effect due to the side gate only and how it may be affected by another side gate as shown in Fig. 3a and b. An ambipolar field effect is observed by sweeping the side gate voltage only. In Fig. 3a, all of the curves show clear side gate "Dirac points". For SG1, $V_{SDP1}$ ~ 70V, which is lower than $V_{SDP2}$ ~90V. This difference may be due to the better capacitive coupling between SG1 and graphene or the charge inhomogeneity in the graphene.[27] When both side gates are used simultaneously, the "joint" field effect has a further reduced $V_{SDP}$ ~50V. Furthermore, we find that the side-gated field effect due to SG1 can be continuously tuned by applying a voltage to SG2 ($V_{sg2}$) as shown in Fig. 3b. Increasing $V_{sg2}$ from -80V to 80V tunes the field effect due to SG1 from p-type behavior (R increases with increasing $V_{sg2}$ within the measurement range) to ambipolar, with $V_{SDP1}$ decreasing from positive to 0V and even to negative (Fig. 3b). Similar results are also obtained when we sweep $V_{sg2}$ at various $V_{sg1}$. The on/off resistance modulation ratio of side-gated field effect can reach ~2 and become comparable to that of back gated field effect (Fig. 1c). The limited on/off ratio from side gated graphene FET may be related to the charge inhomogeneity[23, 28] induced by the non-uniform electric field from the side gate.

We have also observed that the side-gated field effect is sensitive to the electrical grounding of the back gate. This is demonstrated in Fig. 3c, with measurements performed on a device "2" with only one local metal side gate but otherwise similar to device "1". It can be seen that the side-gated field effect with the back gate floating is much stronger than the case with the back gate grounded. We have performed finite element (COMSOL Multiphysics 3.5a) simulations to calculate the spatial electric field profile with various gate configurations. Figure 3d shows the calculated electric field strength at a representative point above graphene as a function of $V_{sg}$ for the two different back gate conditions. It can be seen that the electric field at graphene is stronger with a floating back gate (than a grounded back gate), leading to the stronger field effect observed (Fig. 3c).

In summary, we have demonstrated metal-side-gated ambipolar graphene field effect transistors, fabricated in a one-step process without any gate dielectric deposition or graphene etching. The local metal side gates show promising ability to tune the field effect in graphene and can be used to control individual graphene nanodevices, with many potential applications in carbon-based electronics.

We thank Miller Family Endowment, Midwest Institute for Nanoelectronics Discovery (MIND), Indiana Economic Development Corporation (IEDC), NSF (ECCS-0833689), DHS and DTRA for partial support of this research.

**References**:

[1] A. K. Geim and K. S. Novoselov, Nature Mater. **6**, 183 (2007).

[2] A. K. Geim, Science **324**, 1530 (2009).

[3] C. Berger, Z. Song, X. Li, X. Wu, N. Brown, C. Naud, D. Mayou, T. Li, J. Hass, A. N. Marchenkov, E. H. Conrad, P. N. First, and W. A. de Heer, Science **312**, 1191 (2006).

[4] K. S. Novoselov, A. K. Geim, S. V. Morozov, D. Jiang, M. I. Katsnelson, I. V. Grigorieva, S. V. Dubonos, and A. A. Firsov, Nature (London) **438**, 197 (2005).

[5] Y. Zhang, Y.W. Tan, H. L. Stormer, and P. Kim, Nature (London) **438**, 201 (2005).

[6] K. I. Bolotin, F. Ghahari, M. D. Shulman, H. L. Stormer, and P. Kim, Nature 462, 196 (2009).

[7] X. Du, I. Skachko, F. Duerr, A. Luican, and E. Y. Andrei, Nature **462**, 192 (2009).

[8] F. Schedin, A. K. Geim, S. V. Morozov, D. Jiang, E. H. Hill, P. Blake, and K. S. Novoselov, Nature Mater. **6**, 652 (2007).

[9] S. Pisana, P. M. Braganca, E. E. Marinero, and B. A Gurney, Nano Lett. **10**, 341 (2010).

[10] H. Wang, D. Nezich, J. Kong, and T. Palacios, IEEE Electron Device Lett. 30, 547 (2009).

[11] M. Y. Han, B. Oezyilmaz, Y. Zhang, and P. Kim, Phys. Rev. Lett. **98**, 206805 (2007).

[12] Z. Chen, Y. -M. Lin, M. J. Rooks, and P. Avouris, Physica E **40**, 228 (2007).

[13] M. C. Lemme, T. J. Echtermeyer, M. Baus, and H. Kurz, IEEE Electron Device Lett. **28**, 283 (2007).

[14]J. R. Williams, L. DiCarlo, and C. M. Marcus, Science **317**, 638 (2007).

[15]F. Molitor, J. Guttinger, C. Stampfer, D. Graf, T. Ihn, and K. Ensslin, Phys. Rev. B **76**, 245426 (2007).

[16]B. Huard, J. A. Sulpizio, N. Stander, K. Todd, B. Yang, and D. Goldhaber-Gordon, Phys. Rev. Lett. **98**, 236803 (2007).

[17]B. Ozyilmaz, P. Jarillo-Herrero, D. Efetov, D. A. Abanin, L. S. Levitov, and P. Kim, Phys. Rev. Lett. **99**, 166804 (2007).

[18]X. Li, X. Wu, M. Sprinkle, F. Ming, M. Ruan, Y. Hu, C. Berger, and W. A. de Heer, Phys. Status Solidi A **207**, 286 (2010).

[19]K. Zou, X. Hong, D. Keefer, and J. Zhu, arXiv: 0912.1378v2 (2009).

[20]D. B. Farmer, H. -Y Chiu, Y. -M Lin, K. A. Jenkins, F. Xia, and P. Avouris, Nano Lett. **9**, 4474 (2009).

[21]T. Shen, J. J. Gu, M. Xu, Y. Q. Wu, M. L. Bolen, M. A. Capano, L. W. Engel, and P. D. Ye, Appl. Phys. Lett. **95**, 172105 (2009).

[22]G. Liu, J. Velasco, W. Bao, and C.N. Lau, Appl. Phys. Lett. **92**, 203103 (2008).

[23]J. S. Moon, D. Curtis, S. Bui, M. Hu, D. K. Gaskill, J. L. Tedesco, P. Asbeck, G. G. Jernigan, B. L. VanMil, R. L. Myers-Ward, C. R. Eddy, Jr., P. M. Campbell, and X. Weng, IEEE Electron Device Lett. **31**, 260 (2010).

[24]D. Graf, F. Molitor, T. Ihn, and K. Ensslin, Phys Rev. B **75** 245429 (2007).

[25]K. S. Novoselov, A. K. Geim, S. V. Morozov, D. Jiang, Y. Zhang, S. V. Dubonos, I. V. Grigorieva, and A. A. Firsov, Science **306** 666 (2004).

[26]A. C. Ferrari, Solid State Commun. **143**, 47 (2007).

[27]P. Blake, R. Wang, S. V. Morozov, F. Schedin, L. A. Ponomarenko, A. A. Zhukov, R. R. Nair, I. V. Grigorieva, K. S. Novoselov, and A. K. Geim, Solid State Commun. **149**, 1068 (2009).

[28]I. Meric, M. Y. Han, A. F. Young, B. Ozyilmaz, P. Kim, and K. L. Sheppard, Nat. Nanotechnol. **3**, 654 (2008).

**Figure captions:**

FIG. 1. (a) AFM image of a graphene FET (device "1") with two local metal side gates ("SG1" and "SG2"). The dash line marks the edge of graphene, separated by ~370 nm from the side gates for this device. (b) The 3D schematic of the device and corresponding circuits used in the measurements. (c) The global field effect of device "1" by sweeping the back gate voltage. The FET mobility of this device is ~ 3500 $cm^2$/Vs.

FIG. 2. (a) Resistance (R) of the graphene device as a function of side gate voltage ($V_{sg}$) at various back gate voltages ($V_{bg}$). (b) R as a function of $V_{bg}$ at various $V_{sg}$. (c) Two-dimensional (2D) color plot of R as a function of $V_{bg}$ and $V_{sg}$. Curves in (a) and (b) correspond to vertical and horizontal line cuts in the color plot. All data measured in device "1" using SG1 as the side gate (Fig.1).

FIG. 3. (a) Field effect controlled by one or two side gates (device "1"). (b) R of device "1" as a function of $V_{sg1}$ at various $V_{sg2}$. (c) The side gated field effect measured in another device "2" (similar to device "1", but with one side gate only) with back gate grounded or floating respectively. (d) Calculated electric field strength (by COMSOL, for a device structure similar to that of device "2") at a representative point (1 nm above graphene and 250 nm from the left edge) as a function of $V_{sg}$ at different back gate conditions.

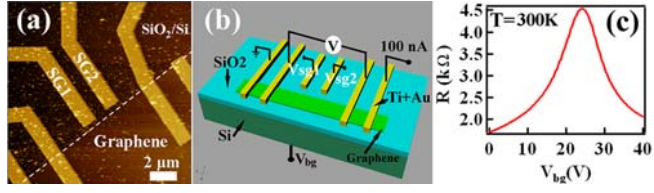

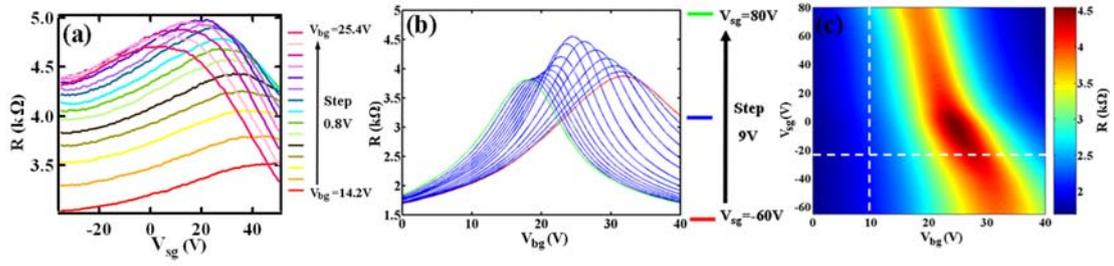

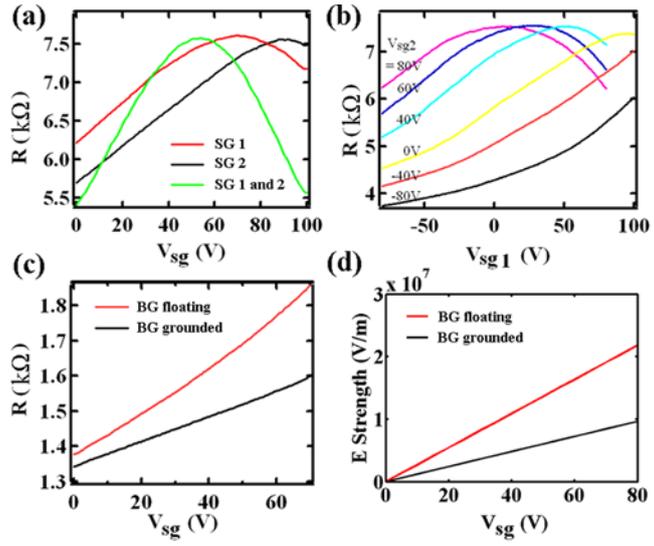
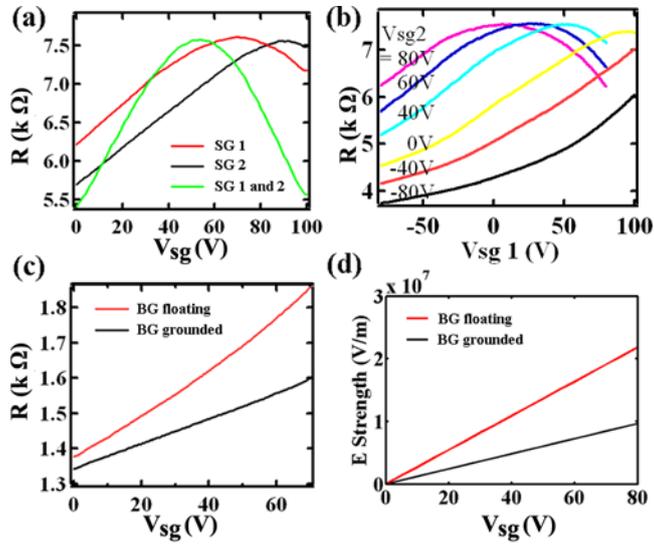